\documentclass{llncs}

\usepackage{amssymb}
\usepackage{amsmath}
\usepackage{graphicx}
\usepackage{color}

\usepackage[OT4]{fontenc}
\usepackage[cp1250]{inputenc}

\newcommand{\mket}[1]{| #1 \rangle}
\newcommand{\mbra}[1]{\langle #1 |}

\newcommand{\keywords}[1]{\par\addvspace\baselineskip\noindent\keywordname\enspace\ignorespaces#1}

\begin{document}


\mainmatter

\title{Transfers of entanglement qudit states in quantum networks}
\authorrunning{Marek Sawerwain \and Joanna Wiśniewska}
\tocauthor{Marek Sawerwain and Joanna Wiśniewska}

\author{Marek Sawerwain\inst{1} \and Joanna Wi\'sniewska\inst{2}}
\institute{Institute of Control \& Computation Engineering \\
University of Zielona G\'ora, ul. Licealna 9, Zielona G\'ora 65-417, Poland \\
\email{M.Sawerwain@issi.uz.zgora.pl}
\and
Institute of Information Systems, Faculty of Cybernetics, \\
Military University of Technology, ul. Kaliskiego 2, 00-908 Warsaw, Poland \\
\email{jwisniewska@wat.edu.pl}
}

\maketitle

\begin{abstract}
The issue of quantum states' transfer -- in particular, for so-called Perfect State Transfer (PST) -- in the networks represented by the spin chains seems to be one of the major concerns in quantum computing. Especially, in the context of future communication methods that can be used in broadly defined computer science. The chapter presents a definition of Hamiltonian describing the dynamics of quantum data transfer in one-dimensional spin chain, which is able to transfer the state of unknown qudits. The main part of the chapter is the discussion about possibility of entangled states' perfect transfer, in particular, for the generalized Bell states for qudits. One of the sections also contains the results of numerical experiments for the transmission of quantum entangled state in a noisy quantum channel.
\end{abstract}

\section{Introduction} \label{lbl:sec:introduction}

The problem of Perfect State Transfer (PST), raised i.a. in \cite{Bose2003}, generates a very important area of quantum computing, which is naturally combined with the context of information transfer in quantum channels. More information on PST can be found in \cite{Bose2007} and \cite{Kay2010}.

A quantum state's transfer from the specified start position to another position is called a perfect state transfer, if the Fidelity value of the initial  and the final quantum state is equal to one. The possibility of perfect transmission
not only in XX-like and XY-like spin chains is shown in \cite{Albanese2004}, \cite{Karbach2005}, \cite{WuLA} and in previously quoted \cite{Bose2007}, \cite{Kay2010}. The individual chain's elements are inhomogeneously coupled. The spin chains with homogeneous coupling provide the transfer only for short chains -- containing three or four elements. It is important that the transfer does not require any additional intervention except the influence of dynamics described by the corresponding Hamiltonian. The techniques of spin construction for PST are also significant -- example in \cite{Vinet2012}. 

Naturally, the main task of perfect state transfer protocols is addressed to transfer quantum state in qubits/qudits 1D chains, and also in more  complex quantum networks, e.g. 2D grid or graphs. However, the use of chain to transfer entanglement states \cite{Horodecki2009} is also possible, but at least two chains must be used -- see paper \cite{Christandl2005}.

It should be also mentioned, that the perfect state transfer (PST) systems are mainly used to construct other primitive protocols e.g. GHZ and W  states preparation, general entanglement generation, state of system initialisation, signal amplification in measurements, realisation of quantum  walks, realisation of universal quantum computation.

The main objective of this chapter is to verify whether the entangled state can be also transferred to a higher-dimensional space with use of spin chains with suitably chosen dynamics.

The content of this chapter is as follows: in part (\ref{lbl:sec:generators:of:lie:algebra}) the definition of Lie algebra's generator was cited  and it is combined with the Hamiltonian definition, which is responsible for sending a quantum state by the spin chain. The spin chain described in introduction of publication \cite{Sawerwain2012} is briefly characterised in (\ref{lbl:definition:of:hamiltonian:for:qudits}). The definition of transferred entangled states and the construction of a spin chain is presented in (\ref{lbl:sec:transfers:entanglement:states}). In subchapter (\ref{lbl:transfers:of:entanglement:states:in:noisy:channels}) are shown: numerical simulations of entangled states' transfer and exemplary simulations of transfer process in an environment where noise is present. The summary and short term objectives are outlined in  (\ref{lbl:sec:conslusions}). The chapter's last section consists of acknowledgments
and a list of cited literature.

\section{Generators of Lie algebra} \label{lbl:sec:generators:of:lie:algebra}

In proposed definition of the XY-like Hamiltonian for qudits' chain (given in the section (\ref{lbl:sec:transfers:entanglement:states}) of this chapter), Lie algebra's generators for a group $SU(d)$, where $d \geq 2$, will be used to define a suitable operator which is responsible for transfer dynamics. For clarity, the following well known set construction procedure of $SU(d)$ generators will be recalled: in the first step, a set of projectors is defined
\begin{equation}
(P^{k,j})_{\upsilon, \mu} = \mket{k}\mbra{j} = \delta_{\upsilon, j} \delta_{\mu, k},  \;\;\; 1 \leq \upsilon, \mu \leq d .
\end{equation}
The first suite of $d(d-1)$ operators from the group $SU(d)$ is specified as
\begin{equation}
\Theta^{k,j}  =  P^{k,j} + P^{j,k}, \;\;\; \beta^{k,j}  = -i (P^{k,j} - P^{j,k}),
\label{MSawe:JWisn:CN2013:QUDITPST:lbl:eqn:theta:beta:operators}
\end{equation}
and $1 \leq k < j \leq d$.

The remaining $(d-1)$ generators are defined in the following way
\begin{equation}
\eta^{r,r} = \sqrt{\frac{2}{r(r+1)}} \left[ \left( \sum^{r}_{j=1} P^{j,j} \right) - r P^{r+1,r+1} \right],
\end{equation}
and $1 \leq r \leq (d-1)$. Finally, the $d^2-1$ operators belonging to the $SU(d)$ group can be obtained.

\begin{remark}
For $d=2$ obtained suite of $SU(d)$ operators is the set of Pauli operators:
\begin{equation}
\begin{array}{ccc}
  \sigma_x = X = \left(%
\begin{array}{cc}
  0 & 1 \\
  1 & 0 \\
\end{array}%
\right),
&
  \sigma_y = Y =\left(%
\begin{array}{cc}
 0  & -i \\
 i  &  0 \\
\end{array}%
\right),
&
  \sigma_z = Z =\left(%
\begin{array}{cc}
 1 &  0 \\
 0 & -1 \\
\end{array}%
\right)
\end{array},
\label{MSawe:QUDITPST:lbl:eqn:pauli:operator}
\end{equation}
while for $d=3$ the set of Gell-Mann operators $\lambda_i$ will be obtained:
\begin{equation}
\begin{array}{lll}
\lambda_1 = \Theta^{1,2} = \left(
\begin{array}{ccc}
 0 & 1 & 0 \\
 1 & 0 & 0 \\
 0 & 0 & 0
\end{array}
\right), &
\lambda_2 = \beta^{1,2} = \left(
\begin{array}{ccc}
 0 & -i & 0 \\
 i & 0 & 0 \\
 0 & 0 & 0
\end{array}
\right), &
\lambda_3 = \eta^{1,1} =  \left(
\begin{array}{ccc}
 1 & 0 & 0  \\
 0 & -1 & 0 \\
 0 & 0 & 0
\end{array}
\right) \\
\lambda_4 = \Theta^{1,3} = \left(
\begin{array}{ccc}
 0 & 0 & 1 \\
 0 & 0 & 0 \\
 1 & 0 & 0
\end{array}
\right), &
\lambda_5 = \beta^{1,3} = \left(
\begin{array}{ccc}
 0 & 0 & -i \\
 0 & 0 & 0 \\
 i & 0 & 0
\end{array}
\right), &
\lambda_6 = \Theta^{2,3} = \left(
\begin{array}{ccc}
 0 & 0 & 0 \\
 0 & 0 & 1 \\
 0 & 1 & 0
\end{array}
\right) \\
\lambda_7 = \beta^{2,3} = \left(
\begin{array}{ccc}
 0 & 0 & 0 \\
 0 & 0 & -i \\
 0 & i & 0
\end{array}
\right) &
\lambda_8 = \eta^{2,2} = \frac{1}{\sqrt{3}} \left(
\begin{array}{ccc}
 1 & 0 & 0 \\
 0 & 1 & 0 \\
 0 & 0 & -2
\end{array}
\right) & \\
\end{array} .
\label{MSawe:QUDITPST:lbl:gell:mann:matrices}
\end{equation}
To define the XY-like dynamics all listed above operators are not necessary -- only operators $\Theta^{k,j}$ and $\beta^{k,j}$ according to equation (\ref{MSawe:JWisn:CN2013:QUDITPST:lbl:eqn:theta:beta:operators}) are used.
\end{remark}

\section{Definition of Hamiltonian for qudits} \label{lbl:definition:of:hamiltonian:for:qudits}

In this chapter the following Hamiltonian $H^{d}_{XY}$ (firstly presented in \cite{Sawerwain2012}) is used to realise the perfect transfer of quantum information in qudits chains. It is also claimed that each qudit has the same level and $d \geq 2$:
\begin{equation}
H^{{XY}_{d}} = \sum_{(i,i+1) \in \mathcal{L}(G)} \frac{J_i}{2} \left( \Theta^{k,j}_{(i)} \Theta^{k,j}_{(i+1)} + \beta^{k,j}_{(i)} \beta^{k,j}_{(i+1)} \right), 
\label{lbl:eqn:qudit:pst:hamiltonian}
\end{equation}
where $J_i$ is defined as follows: $J_i = \frac{\sqrt{i(N-i)}}{2}$ for $1 \leq k < j < d$ and $\Theta^{k,j}_{(i)}$, $\beta^{k,j}_{(i)}$ are $SU(d)$ group operators defined by (\ref{MSawe:JWisn:CN2013:QUDITPST:lbl:eqn:theta:beta:operators}) applied to the $(i)$-th and $(i+1)$-th qudit. The Hamiltonian (\ref{lbl:eqn:qudit:pst:hamiltonian}) will be also called the transfer Hamiltonian.  

It is not hard to show that
\begin{equation}
[H^{{XY}_{d}}, \sum_{i=1}^{N} \eta^{r,r}_{(i)}] = 0
\end{equation}
for $1 \leq r \leq (d-1)$, as in the definition of the XY-like Hamiltonian for qubits.

\section{Transfer's example for entangled states} \label{lbl:sec:transfers:entanglement:states}

The entangled states -- so-called Bell states -- for two qubits are:

\begin{equation}
\mket{\psi^{\pm}} = \frac{1}{\sqrt{2}}  \left( \mket{00} \pm \mket{11} \right), \;\;\; \mket{\phi^{\pm}} = \frac{1}{\sqrt{2}}  \left( \mket{01} \pm \mket{10} \right)
\end{equation}

It means there are four Bell states for two-qubit system (respecting the sign).

For qudits with $d$ levels, more complex units than qubits, the equivalent of state $\mket{\psi^{\pm}}$ is
\begin{equation}
\mket{\psi^{\pm}} = \frac{1}{\sqrt{d}}  \left( \mket{00} \pm \mket{nn} \right),
\end{equation}
where $n = d - 1$.

The definition of generalised Bell states for qudits with freedom level $d$ is as follows
\begin{equation}
\mket{ \psi_{pq} } =  \frac{1}{\sqrt{d}} \sum_{j=0}^{d-1} e^{2 \pi i j p / d} \mket{j} \mket{(j+q) \, \textrm{mod} \, d} \;\;\; 0 \leq p, q \leq d - 1,
\end{equation}
where $i$ represents an imaginary unit. Generally exist $d^2$ Bell states for two d-level qudits.

It is convenient to express the last equation as a circuit of quantum gates for generalised EPR pair.
\begin{equation}
\mket{\psi^d_{pq}} = {(I_d \otimes X_d)}^q \cdot {(H_d \otimes I_d)} \cdot {(Z_d \otimes I_d)}^p \cdot \mathrm{CNOT_d} \cdot \mket{00}  \label{lbl:dLevel:Bell:State:gate}
\end{equation}
\begin{remark}
The above form of EPR pair is also used in a process of transferred state's correction -- for the transfer of maximally entangled qudits, the transferred qudit is still maximally entangled, but the values of amplitudes usually represent other maximally entangled state. Other examples of the use of entanglement states in transfer procotol can be found i.e. \cite{WangMingMing2012}.
\end{remark}


For unknown pure state of one qudit
\begin{equation}
\mket{\psi} = \alpha_0\mket{0} + \alpha_1\mket{1} + \ldots + \alpha_{d-1}\mket{d-1} \;\;\; \mathrm{and} \;\;\; \sum_{i=0}^{d-1}{|\alpha_i|}^2  = 1, \; \mathrm{where} \; \alpha_i,  \in \mathbb{C} ,
\label{lbl:eqn:qudit:pure:state}
\end{equation}
the transfer process (or transfer protocol) in a one-dimensional chain of $n$~qudits is expressed as a~transformation of the state~$\mket{\Psi_{\mathrm{in}}}$ into the state~$\mket{\Psi_{\mathrm{out}}}$:
\begin{equation}
\mket{\Psi_{\mathrm{in}}} = \mket{\psi}\mket{\underbrace{000 \ldots 0}_{n-1}} \;\;\; {\Longrightarrow} \;\;\; \mket{\Psi_{\mathrm{out}}} = \mket{\underbrace{000 \ldots 0}_{n-1}}\mket{\psi} .
\label{MSawe:JWisn:CN2013:lbl:eqn:task:transfer:equation}
\end{equation}

\begin{figure}
\begin{tabular}{c}
\includegraphics[height=5.00cm]{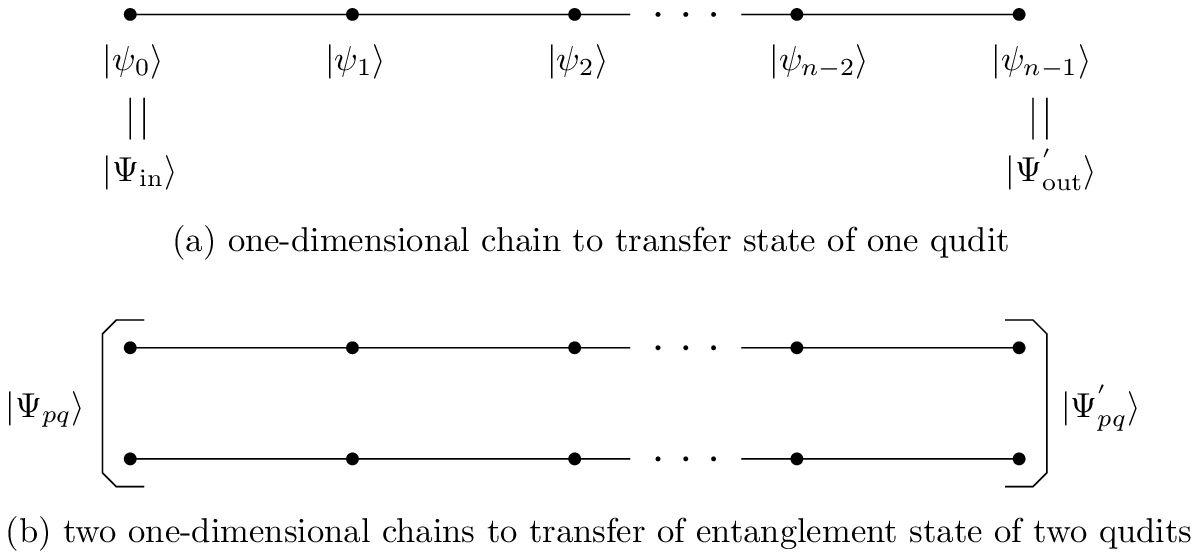} 
\end{tabular}
\centering
\caption{The realisation of information transfer in qudits chains for single (a) and entanglement states (b), the interactions between qudits are performed only between neighbourhood qudits}
\label{MSawe:JWisn:CN2013:lbl:fig:transfer:information:in:qudits:chains}
\end{figure}

If the transfer is performed on entangled state of two qudits, the transfer protocol realises the transmission of both qudits' quantum state. Naturally, the entanglement -- as the state's feature -- have to be also transferred. The state $\mket{\Psi_{\mathrm{out}}}$ corresponds to $\mket{\Psi_{\mathrm{in}}}$ according to selected value of Fidelity:
\begin{equation}
\mket{\Psi_{\mathrm{in}}} = \mket{\psi_{pq}}\mket{\underbrace{000 \ldots 0}_{n-2}} \;\;\; {\Longrightarrow} \;\;\; \mket{\Psi_{\mathrm{out}}} = \mket{\underbrace{000 \ldots 0}_{n-2}} \mket{\psi_{pq}} .
\label{MSawe:JWisn:CN2013:lbl:eqn:task:transfer:equation:for:qudits}
\end{equation}
where $\mket{\psi_{pq}}$ is a two-qudit system. The spin chains for transferring qudits' states are briefly shown at the Figure (\ref{MSawe:JWisn:CN2013:lbl:fig:transfer:information:in:qudits:chains}).

En exemplary transfer of Bell state $\mket{\psi^{+}}$ with use of Hamiltonian (equation (\ref{lbl:eqn:qudit:pst:hamiltonian})) for four-qubit spin chain is a perfect state transfer. It means there is no addictional correction needed. After the transfer process the obtained state is $\mket{\psi^{+}}$ Bell state. Generally, during the transfer the chain's state is:
\begin{equation}
\mket{\Psi_t} = \alpha_0 \mket{0000} + \alpha_1 \mket{0011} + \alpha_2 \mket{0110} + \alpha_3 \mket{1001} + \alpha_4 \mket{1100},
\end{equation}
and the final chain's state may be expressed as:
\begin{equation}
\mket{\Psi} = \alpha_0 \mket{0000} + \alpha_1 \mket{0011}
\end{equation}
For Bell state $\mket{\phi^{+}}$ its transfer is a perfect state transfer in terms of Fidelity value. However, it has to be taken into account that the amplitudes' values are multiplied by imaginary unit~$i$.

\begin{remark}
Naturally, the Hamiltonian (\ref{lbl:eqn:qudit:pst:hamiltonian}) for qubits is XY-like Hamiltonian \cite{Bose2003}, \cite{Christandl2005}.
\end{remark}

It is easy to calculate the index of each node for a $n$-node spin chain. Of course, these values depend on transferred state. For the transmission of Bell state $\mket{\psi^{+}}$ -- using binary codding and realising bit shift operations -- the states are used where bit value $(11)_{2}$ is shifted through all chain's nodes:
\begin{equation}
\begin{array}{c}
\mket{11000 \ldots 000} \\
\mket{01100 \ldots 000} \\
\mket{00110 \ldots 000} \\
\mket{00011 \ldots 000} \\
\mket{00000 \ldots 110} \\
\mket{00000 \ldots 011}
\end{array}
\end{equation}
the similar action for $(1001)_{2}$:
\begin{equation}
\begin{array}{c}
\mket{1001000 \ldots 00000} \\
\mket{0100100 \ldots 00000} \\
\mket{0010010 \ldots 00000} \\
\mket{0001001 \ldots 00000} \\
\mket{0000000 \ldots 10010} \\
\mket{0000000 \ldots 01001} 
\end{array}
\end{equation}
where the complement states are:
\begin{displaymath}
\mket{1000 \ldots 001} \;\;\; \textrm{oraz} \;\;\; \mket{00000 \ldots 000}.
\end{displaymath}

The entangled states may also be changed (according to the length of spin chain), if the qudits are transferred with use of the Hamiltonian (\ref{lbl:eqn:qudit:pst:hamiltonian}). Mentioned change respects to the phase form and to the entangled state itself. However, the obtained state is still maximally entangled.


Just like for $\mket{\psi^{+}}$ the transfer of state:
\begin{equation}
\mket{\psi} = \frac{1}{\sqrt{3}}\left( \mket{00} + \mket{11} + \mket{22} \right)
\end{equation}
is perfect (PST) and no additional conversion is needed. Just like before, adding the fifth qudit to the spin chain will change the state after the transfer as follows:
\begin{equation}
\mket{\psi} = \frac{1}{\sqrt{3}}\left( \mket{00} - \mket{11} - \mket{22} \right)
\end{equation}
For the transfer of state:
\begin{equation}
\mket{\psi} = \frac{1}{\sqrt{3}}\left( \mket{01} + \mket{12} + \mket{20} \right)
\end{equation}
the state after transfer is:
\begin{equation}
\mket{\psi'} = \frac{1}{\sqrt{3}}\left( \mket{02} + \mket{10} - \mket{12} \right)
\end{equation}
The double use of gate $X$ is needed on the second qudit of state $\mket{\psi'}$ to obtain the entangled state. Of course, this local operation will not change the level of entanglement.




Generalising, the length of spin chain for transferring entangled states affects on amplitudes' phase shifts (it is important if the number of nodes is odd or even). The transmission of state $\mket{\psi^{+}}$ in spin chain with five nodes causes that the state $\mket{\psi^{-}}$ will be obtained. Naturally, the state will be still maximally entangled.


\begin{remark}
The issue of phase shift or amplitudes' permutation in maximally entangled states causes no problems, because there are deterministic procedures of Bell states detection both for qubits \cite{GuptaPanigrahi} and qudits \cite{GuptaPathakSrikanthPanigrah} in generalised Bell states. Using the circuits described in cited papers it is possible to undoubtedly identify, without any damage on EPR pair, the quantum state after the process of transfer.
\end{remark}

The package QCS (Quantum Computing Simulator) -- developed at the University of Zielona Góra -- was used in the experiment for the numerical simulation of spin chain's behaviour. The program's main loop for transfer's simulation is briefly presented at Fig. (\ref{MSawe:JWisn:CN2013:lbl:fig:qcs:script:for:transfer:information:in:qudits:chains}).

\begin{figure}
\begin{center}
\begin{tabular}{l|l}
\hline\hline
~~~~ import qcs										& ~~~~ \# simulation loop \\	
													& ~~~~ s=0 \\	
~~~~ \# five qudits with d = 3						& ~~~~ while $i < 8$: \\	
~~~~ q = qcs.QuantumReg(5,3) 						& ~~~~ ~~~~ ... other operations \\	
~~~~ q.Reset() 										& ~~~~ ~~~~ q.ApplyOperator( op ) \\	
													& ~~~~ ~~~~ ... other operation \\	
~~~~ \# transfer operator							& ~~~~ ~~~~ ... e.g. noise introduction \\
~~~~ op = q.XYTranHamiltonian(						& \\
~~~~ ~~~~ ~~~~ \_fromqudit=0,						& \\
~~~~ ~~~~ ~~~~ \_toqudit=4,							& ~~~~ \# display state of \\
~~~~ ~~~~ ~~~~ \_step = 8)							& ~~~~ \# quantum register \\	
													& ~~~~ q.Pr() \\	
~~~~ \# create generalised Bell state				& \\	
~~~~ \# e.g. $\mket{00} + \mket{11} + \mket{22}$	& \\	
~~~~ \# at qudits zero and one 						& \\	
~~~~ q.SetGBellState(0,1,0,0)						& \\ \hline\hline
\end{tabular}
\end{center}
\caption{A Python script using QCS package to simulate entangled states' transfer in qudit chain. The process of transfer is realised by eight simulation steps}
\label{MSawe:JWisn:CN2013:lbl:fig:qcs:script:for:transfer:information:in:qudits:chains}
\end{figure}

\section{Transfers of entangled states in noisy channels} \label{lbl:transfers:of:entanglement:states:in:noisy:channels}

\begin{figure}[!ht]
\begin{tabular}{c}
\includegraphics[height=5.00cm]{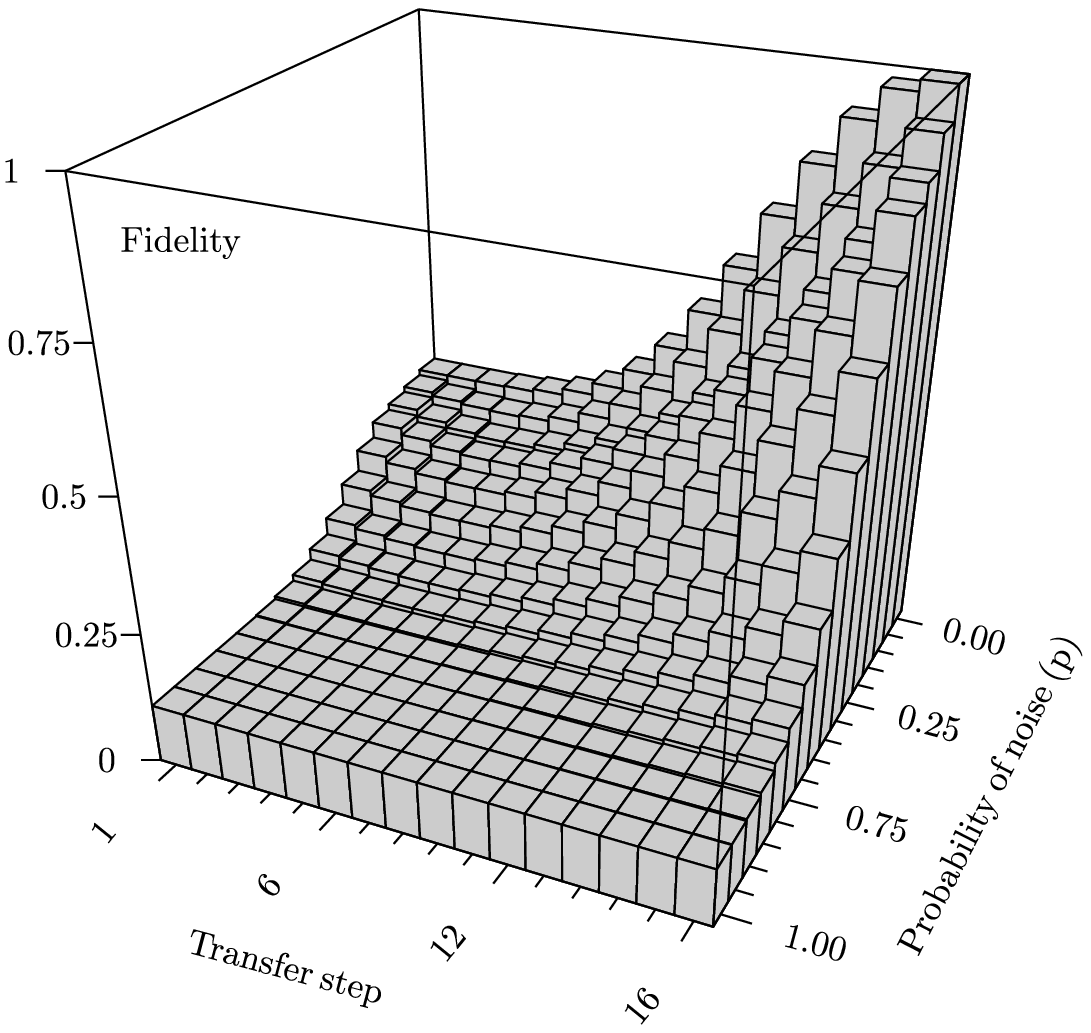}  \includegraphics[height=5.00cm]{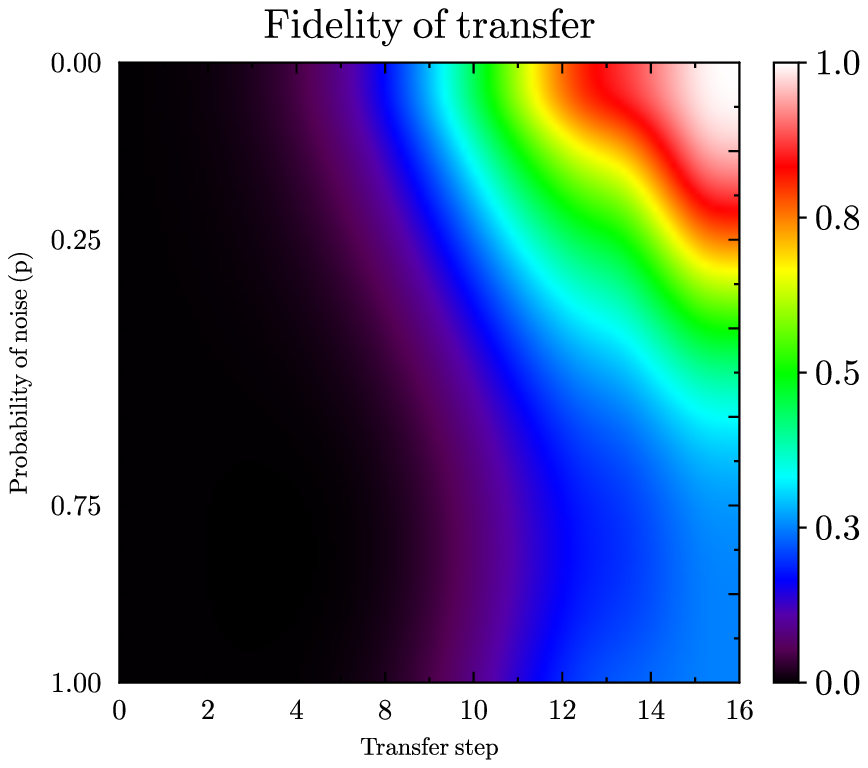} 
\end{tabular}
\centering
\caption{The diagram of Fidelity value for maximally entangled state transfer in five qutrits' chain. The $p$ parameter describes the phase distortion for each qudit. The transfer process consists of sixteen discrete steps}
\label{MSawe:JWisn:CN2013:lbl:fig:fidelity:of:transfer:information:in:noisy:qudits:chains}
\end{figure}

The research in the field of quantum circuits, where noise is present, is a very important issue of quantum computing because of the decoherence phenomenon -- the paper \cite{Gawron2012} shows the impact of noise in quantum channels on: amplitude-damping, phase-damping and bit-flip in Grover's algorithm for database search.
\begin{figure}[!ht]
\begin{tabular}{c}
\includegraphics[height=2.75cm]{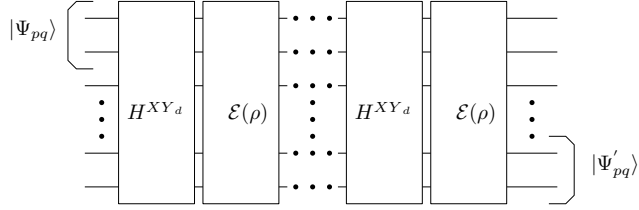} 
\end{tabular}
\centering
\caption{The transfer of entangled state in a noisy channel implemented as a quantum circuit. The first part of the transfer is the operation $X^{XY_d}$. The second part of the circuit represents the noise. The transfer process consists of discrete stages, so the stages of operation $X^{XY_d}$ and noise operation must be repeated}
\label{MSawe:JWisn:CN2013:lbl:fig:transfer:information:in:noisy:qudits:chains}
\end{figure}
In this chapter only the phase-dumping influence on qudit chain for entangled state transfer will be presented.
However, in the case of qudit the phase-damping operation does not have a unique representation. The model discussed in the works \cite{Fukuda} is an example of the phase-damping operation and it will be used in this chapter:
\begin{equation}
\mathcal{E}(\rho) = \sum_{i=0}^{d-1} E_i \rho E^{\dagger}_i, \;\;\; E_i = \sqrt{ {{d-1} \choose {i}} {\left( {1-p} \over 2 \right)}^i {\left( {1+p} \over 2 \right)}^{d-1-i} } Z^i ,
\label{MSawe:JWisn:CN2013:lbl:eq:phase:damping:qudit}
\end{equation}
where $0 \leq p \leq 1$. 
\begin{remark}
It should be noted that expression (\ref{MSawe:JWisn:CN2013:lbl:eq:phase:damping:qudit}) can be regarded as a~special case of Weyl's channel \cite{Fukuda}:
\begin{equation}
\mathcal{E}(\rho) = \sum\limits_{m,n = 0}^{d-1} \pi_{m,n} (Z^{n}X^{m}) \rho {(X^{m} Z^{n})}^{\dagger} ,
\end{equation}
where elements of the matrix $\pi$ satisfy the following conditions: $0 \leq \pi_{m,n} \leq 1$ and $\sum_{m,n=0}^{d-1} \pi_{m,n} = 1$. The operators $Z$ and $X$ are generalised Pauli matrices for the sign changing and negation operations on qudits.
\end{remark}
At the Fig. (\ref{MSawe:JWisn:CN2013:lbl:fig:transfer:information:in:noisy:qudits:chains}) the process of transfer with noise-adding operation is shown. The noise type is phase-damping -- presented in equation (\ref{MSawe:JWisn:CN2013:lbl:eq:phase:damping:qudit}). The diagram of Fidelity value for entangled qudit state transfer $\mket{\psi} = 1 / \sqrt{3} \left( \mket{00} + \mket{11} + \mket{22} \right)$, where $d=3$, is shown at the Fig. (\ref{MSawe:JWisn:CN2013:lbl:fig:fidelity:of:transfer:information:in:noisy:qudits:chains}). Increasing the value of $p$ parameter means the reduction of phase-damping impact on transfer process and the value of Fidelity will be raising in accordance with next stages of the process. It should be pointed that the value of Fidelity is insensitive for the change of the sign or probability amplitudes' phase-shift. However, in the case analysed in this chapter, the distortion of transfer process does not change the transferred state, but only distorts the level of entanglement. Of course, the transferred state is not a maximally entangled state. It means that the small noise level, for example $p=0.05$, is acceptable and still provides the high value of Fidelity.

\section{Conclusions} \label{lbl:sec:conslusions}

Naturally, for entangled states transfer the next step is multi-qudit-state transmission. It is probable that this type of operation is possible, although the number of additional local corrections performed on transferred state will be greater than the number of these operations for EPR pairs.

The analytical solutions for spin chains are direct consequence of the correctness of the Hamiltonian (\ref{lbl:eqn:qudit:pst:hamiltonian}). However, their form and the correctness of Hamiltonian's construction is analysed in presently prepared work \cite{SawerwainMInPrep}. It should be also mentioned that the analytical form of spin chain, with strictly specified number of nodes, makes the effective numerical calculations possible, because the number of nodes in qudit chain is much lower than the number of states in quantum register.

Another very important issue is the entangled states engineering, e.g. \cite{Kerr1}, \cite{Kerr2}, \cite{Kerr3}, which may be implemented by modified Hamiltonian describing transfer's dynamics in spin chains.

\subsubsection*{Acknowledgments}

We would like to thank for useful discussions with the~\textit{Q-INFO} group at the Institute of Control and Computation Engineering (ISSI) of the University of Zielona G\'ora, Poland. We would like also to thank to anonymous referees for useful comments on the preliminary version of this paper. The numerical results were done using the hardware and software available at the ''GPU $\mu$-Lab'' located at the Institute of Control and Computation Engineering of the University of Zielona G\'ora, Poland.

\end{document}